\RequirePackage{fix-cm}
\documentclass{svjour3}                     
\smartqed  
\usepackage{color}
\usepackage{amssymb}
\usepackage{array}
\usepackage{booktabs}
\usepackage{multirow}
\usepackage{subfigure}
\usepackage{graphicx}
\usepackage{dcolumn}
\usepackage{bm}
\usepackage[mathlines]{lineno}

\usepackage{mathptmx}      

\usepackage{latexsym}
\begin{document}

\title{X-ray diffraction line profile analysis of KBr thin films}


\author{R. Rai         \and
        Triloki   \and B. K. Singh 
}


\institute{R. Rai \at
              Department of Physics, Institute of Science, Banaras Hindu University, Varanasi-221005, India.
            \and
           Triloki \at
             Solid State and Structural Chemistry Unit, Indian Institute of Science, Bangaluru-560012, India.
	\and
     B. K. Singh \at
Department of Physics, Institute of Science, Banaras Hindu University, Varanasi-221005, India.
 \email{ bksingh@bhu.ac.in}
}

\date{Received: date / Accepted: date}

\maketitle

\begin{abstract}
In the present work, the microcrystalline characteristics of KBr thin films have been investigated by evaluating the breadth of diffraction peak. The Williamson-Hall, the Size-Strain Plot and the single line Voigt methods are employed to deconvolute the finite crystallite size and microstrain contribution from the broaden X-ray profile. The texture coefficient and dislocation density have been determined along each diffraction peak. Other relevant physical parameters such as stress, Young's modulus and energy density are also estimated using Uniform Stress Deformation and Uniform Deformation Energy Density approximation of Williamson-Hall method. 
\end{abstract}


\section{Introduction}
\label{sec1}

Enormous past and present experiments in the fields of astroparticle and  high energy physics are  based on the detection of ultraviolet photons emitted by charged particles in a wide variety of mediums from atmosphere to interstitial objects. The vital element for these experiments is a competent photodetector and the photoconversion efficiency is  exploited as a scaling parameter for determining the performance of entire detecting device. The application of  photocathode with the high photoejection probability is an effective means of obtaining increased quantum efficiency (QE). Single-photon sensitivity, excellent time resolution, low signal to noise ratio and most important, high QE, are some essential criteria for an ideal photocathode based device. A hypothetical photocathode comprising all these qualities would certainly beneficial for the extensive developments of the different physics frontiers, including ultraviolet astronomy \cite{SUMER,over,PHEBUS}, underground search of Weakly Interacting Massive Particles (WIMP) for dark matter experiments \cite{E. Figueroa}, multiple electron beam lithography \cite{ST Coyle}, Ultrafast Electron Microscope (UEM) \cite{Sergei}, generation of highly instance free electron laser beams \cite{J. Workman}, gaseous photon detectors \cite{D. Mormann}, scintillation detectors \cite{Daisuke}, medical imaging \cite{Wei}, positron emission tomography \cite{F. Garibaldi} etc.

Aalkali halides thin film photocathodes are persistently  employed to enhance the photosensitivity in the extreme ultraviolet (EUV, 10 nm \textless $ \lambda$ \textless 100 nm) and vacuum ultraviolet (VUV, 100 nm \textless $\lambda$ \textless 200 nm) spectral ranges. Owing to high QE and good stability in EUV to VUV wavelength regions, KBr thin films are a feasible alternative for ultraviolet (UV) and soft X-ray photocathode devices ~\cite{OHW}. These photocathodes are also used as a protective layer in visible-sensitive photon detectors ~\cite{A. Breskin1996} and the lower photoconversion efficiency of KBr photocathode in the far ultraviolet (FUV) wavelength region (typically $>$ 160 nm), improved the signal quality by rejecting  the sources of radiation and background near UV wavelength ~\cite{Oswald}. From previous studies, lots of statistics are available on QE of KBr ~\cite{Tremsin,Weidong,Rai} and other alkali halide thin film photocathodes ~\cite{Luna,Nitti,Buzulutskov,BKSingh,Singh1,Shefer}, but very less work has been done on microstructural characterization  of these photocathodes. Although, the thin film photocathode with a better crystal structure and fewer impurities, have  a higher probability that the ejected photoelectron will survive through transport towards the vacuum interface and contribute in the quantum yield. 
Therefore, in this manuscript, we report on a detailed study aimed at elucidating the variation in the microstructural parameters of KBr thin film with increasing film thickness.

XRD line profile analysis is a powerful tool to quantify the microstructural features of a crystalline materials by evaluating the shape and broadening of Bragg's diffraction. The position and breadth of diffraction peak are realized to be affected by the imperfect crystal structure. This deviation from the crystallinity is attributed to the existence of lattice distortion due to the finite crystallite size and the microstrain. In perturbed crystal structure, the most prominent lattice defect producing a heterogeneous microstrains is dislocation. In proximity of dislocation the atoms reside on the equilibrium position and causes surrounding bond length to contract or expand ~\cite{BooK}. The other sources of strain can be triple junctions of grain or subgrain boundaries, contact or sinter stresses, stacking faults, antiphase boundaries, twin boundaries, coherency stresses, external biaxial strain etc \cite{T}. Like dislocation, the strain fields of triple junctions and contact or sinter stresses are usually heterogeneous and extend over many hundreds of unit cell. The stacking faults, twin boundaries and antiphase boundaries are planar defects and generate  homogeneous strain. While, the external biaxial strain produces from the film growth on lattice mismatched substrate ~\cite{Kisielowski}. The influence of each defect type on a peak broadening can be separated on the basis of their characteristic $hkl$ dependence and, to some extent, on the basis of their different profile shapes \cite{TAMÁS UNGÁR}. However, in this article, we only present a quantitative evaluation of size and strain parameters derived from diffracted peak width of KBr films. 

The experimentally observed diffraction profile is a convolution function of instrument response and microstructural effects. Several analytical methods such as Warren-Averbach \cite{Warren}, single line Voigt  \cite{Langford}, Size-Strain Plot \cite{Prince} and Williamson-Hall \cite{Hall}  are often utilized to deconvolute the finite crystallite size and microstrain contribution from a broaden X-ray profile. Among all available decomposition methods, W-H plots are simplified approach related to the dependence of integral width as a function of the diffraction vector to the volume weighted average crystallite sizes and mean square stain. In the original W-H plots, the peak width is assumed to be a monotonic function of the diffraction order, however, in  its modified versions line broadening due to anisotropic strain has also been taken into account with the presumption of peak shape. In the current work, the W-H method is employed to extract the individual contribution from the size and strain induced peak broadening and obtained results are compared with Size-Strain Plot and single line Voigt methods.
 
\section{Experimental Details}
 KBr films are grown in a high vacuum (~$10^{-7}$ Torr) stainless steel chamber of 18" diameter through resistive vapor evaporation technique.  The atmosphere of the chamber is evacuated by a turbo molecular pump of pumping speed 510 L/s for dry N$_{2}$. At the base pressure $3.5\times10^{-7}$ Torr, the residual contaminants inside the chamber is monitored by a residual gas analyzer (SRS RGA 300). A small amount of KBr powder of 5N purity (Alfa Aesar) has been  placed on a tantalum boat and this boat is kept at a 20 cm distance from a stainless steel substrate. Prior to film preparation, the boat and powders are throughly outgassed up to the temperature close to the melting point of KBr. 
 After several hours of pumping, the abundance of N$_{2}$ gas is observed, while water vapor and other residual gases are present in a very small amount. At this stage KBr powder  is evaporated at the rate of $\sim$3 nm/s.  The deposition rate and films thickness are monitored by a quartz crystal based thickness/rate monitor (Sycon STM-100). Immediately after deposition, the chamber is purged with dry N$_{2}$ in order to ensure the minimum contact with atmospheric air during the sample transfer and prepared films are extracted into a vacuum desiccator which contained fresh silica gel. Further, KBr films are transported to the X-ray diffractometer for the microsturactural  measurements.

The X ray diffraction line profile data is recorded in a continuous scan mode (2$\theta= 10^{o}-85^{o}$) using X'Pert PRO (PANanalytical) diffractometer in the Bragg-Brentano parafocusing  configuration ($\theta/2\theta$ geometry) with CuK$_{\alpha}$ ($\lambda$ = 1.5406~\AA) radiation at room temperature (25$^{o}$C) . The electrons, emitted from the cathode filament are accelerated towards the anode plate (Cu) by applying 40 kV voltage and 30 mA filament current. In X-ray diffractometer Ni foil is used as edge filter to suppress the CuK$_{\beta}$ line. The diffraction beam optics equipped with 0.04 rad solar slit, fixed divergent slit (slit size = 0.8709$^{o}$), 0.100 mm size receiving slit  and a scintillator detector. The samples are scanned with a constant step of 0.02 deg of $2\theta$ and with a constant counting time of 1.5 seconds at each step.  In order to minimize  the instrumental contribution in XRD line broadening, diffractometer has been calibrated with standard silicon (Si) crystal.

\newcommand{\head}[1]{\textnormal{\textbf{#1}}}
\newcommand{\normal}[1]{\multicolumn{1}{l}{#1}}
\section{Results and Discussion} 
\subsection{XRD analysis}
 \begin{figure}[!ht]
  \begin{center}
\includegraphics[scale=0.35]{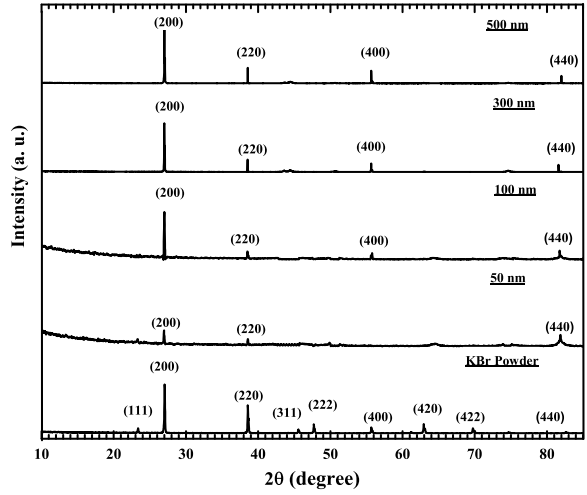} 
 \caption{XRD patterns of KBr powder and of 500 nm, 300 nm, 100 nm, 50 nm thick films.}
     \label{XRD}
    \end{center}
  \end{figure}
The output from the XRD analysis of KBr powder and thin film samples yields a plot of intensity versus diffraction angle as shown in Figure 1. All the detectable XRD peaks are attributed to the face centered cubic (fcc) structure and matched with the standard reference data (JCPDS pdf no.: 730381). The XRD pattern of 100, 300 and 500 nm thick KBr films exhibit the most intense peak at (200) crystallographic plane, followed by  other peaks at (220), (400), (440) planes respectively. In the case of 50 nm film, only three peaks are observed, most intense at (200)  plane and two others  at (220) and (440) plane, while (400) peak excluded from current analysis due to its low count. The structural  parameters correspond to the most intense peak of KBr films are shown in Table 1. 

\begin{table*}[ht]
\renewcommand{\arraystretch}{0.70}
\caption{Structural parameters of KBr films corresponding to (200) crystallographic plane.}
 \vskip .1cm
\centering

\begin{tabular}{c  c  c  c  c}

\hline
 Sample  &2$\theta$           & Lattice constant & Primitive cell         & \% of contraction in
   \\ 
Thickness    & $(degree)$& $a_{exp} (nm)$ & Volume $(nm)^{3}$ & interplaner spacing \\ [1.5ex]

\hline

50 nm & 26.984 &0.660862 &0.072156& 0.3648 \\

100 nm & 26.997& 0.660538 & 0.072059& 0.3156  \\
300 nm&27.000&0.660476&0.072029&0.3062\\
500 nm&27.029&0.659984&0.071868&0.2314\\
\hline
\hline
\end{tabular}
\label{table:nonlin}

\end{table*}

The preferred orientation of films growth can be determined quantitatively by calculating the texture coefficient (TC) along each diffraction planes. The TC$_{(hkl)}$ of each $(hkl)$ plane is evaluated from the XRD data according to the following formula \cite{Barret}
\begin{equation} 
TC_{(hkl)}=\frac{I_{(hkl)}/I_{(hkl)o}}{N^{-1}\sum_{N} I_{(hkl)}/I_{(hkl)o}}
\end{equation}

where $I_{(hkl)}$ is the measured XRD peak intensity; $I_{(hkl)o}$ is the intensity of randomly oriented KBr powder taken from a stranded reference data and  $N$ is the total number of diffractions under consideration. The results, obtained from above equation are tabulated in Table 3. The higher value of the texture coefficient ($TC_{(hkl)}>1$) indicates a higher degree of preferred orientation along a particular plane of the film. The deviation in the texture coefficient from unity for a particular plane attributed to change in atomic densities corresponding to that plane as X-ray intensities are function of atomic structure factors \cite{Cullity}. The value $0<TC_{(hkl)}<1$ represents a decrease in atom density along a particular crystal plane as compared to reference data. The texture analysis reveals that KBr film of thickness 100, 300, 500 nm are highly textured along (200) and (400) plane. However, 50 nm film preferentially grows along (220) and (440) planes. 

\subsection{Crystallite size and Microstrain}
 The XRD pattern of KBr films reveals that all the peaks position are shifted  towards the lower $2\theta$ values with respect to peak positions given in the reference data and with increasing film thickness approaches to the value given for KBr powder. The peak shift towards lower $2\theta$ side indicates that the KBr lattices are in a state of strain. The nature  and upper bound of lattice strain can be evaluated using the following relation:

\begin{equation} 
\frac{\Delta d}{d}   =\frac{d_{exp} - d_{std}}{d_{std}}
\end{equation}

\begin{figure}[!ht]
  \begin{center}
\includegraphics[scale=2.2]{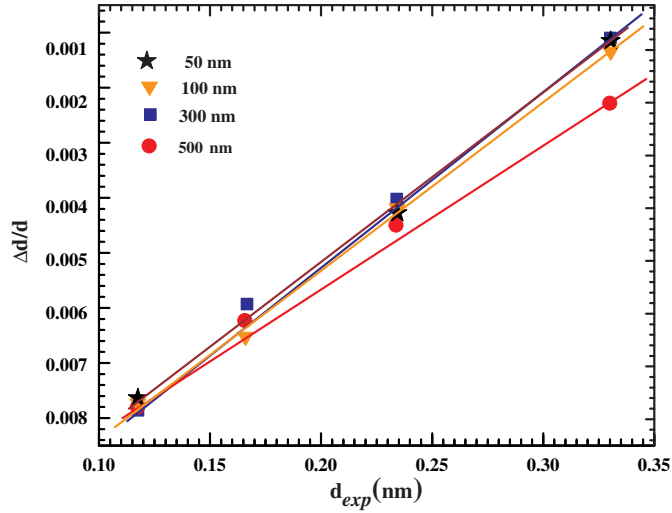}
\caption{The variation of lattice strain (${\Delta d}/{d}$) as a function of $d_{exp}$.}
     \label{AFM_2d_3d}
    \end{center}
  \end{figure}
Where, $d_{exp}$ is the experimentally observed interplaner spacing of KBr films and $d_{std}$ represents the strain- free interplaner spacing as reported in the reference data.  The positive value of ${\Delta d}/{d}$  corresponds the compressive nature of lattice strain field and consequently, this shifted the d-spacing of adjacent crystallographic planes towards higher values. The variation of ${\Delta d}/{d}$ with respect to $d_{exp}$ is shown in Figure 2. However, in the case of KBr films, not only the peak broadening but also the shift in centroid $2\theta$ value is observed, this ascertained that homogeneous as well as heterogeneous strain fields are present in the films and cause lattice distortion. The analysis of Figure 2 ascertains that the influence of heterogeneous strain field is decreased with increasing film thickness as the dislocation density decreases and the atoms try to occupy the equilibrium position in order to minimize the lattice energy. This results in small contraction of interplaner separation and approaches to the value reported for strain free lattices. 

\subsubsection{Scherrer Method:}
Among all the existing analytical techniques applied to investigate the line broadening of X-ray diffraction, the Scherrer method ~\cite{Scherrer} is a most  simplified formulation and therefore still employed to estimate the ``apparent" domain sizes of physical broaden peak profile. This method defines the crystallite size in terms of a mean effective size of the coherently scattering region normal to the reflecting planes~\cite{YU}. The Scherrer relation between crystallite size and integral breadth is given by: 
\begin{equation}
  D= \frac{K \lambda }{ \beta_{D} \cos\theta} \Rightarrow \beta_{D}= \frac{K \lambda }{D \cos\theta}
\end{equation}

where, $D$ is the effective crystallite size normal to the reflecting plane, $K$ is a shape factor (for cubic crystal $K$ = 0.9), $\lambda$ is the wavelength of CuK$_{\alpha}$ radiation, $\beta_{D}$ is the integral width of a particular peak and $\theta$ is the diffraction angle. From equation (3) it is evident that size broadening is independent of order of a reflection. The dislocation density ($\delta$), defined as the length of dislocation lines per unit volume of the crystal, is estimated from the formula \cite{Williamson}

\begin{equation}
  \delta = \frac{1}{D^{2}}
\end{equation}
The smaller value of dislocation density shows the better crystallization of the film along a particular plane. All of these values are reported in Table 3. The second order reflection of all planes shows the larger value of $\delta$ which clearly reveals the deterioration in the
 film crystallinity.

The  strain induced peak broadening resulting from lattice distortion (microstrain) can be expressed by Wilson formula \cite{Stokes}:
\begin{equation}
  \epsilon = {\frac{1}{4}}{\beta_{S}\cot\theta}  \Rightarrow \beta_{S}= 4\epsilon \tan\theta
\end{equation}  
Where, $\epsilon$ is the microstrain and $\beta_{S}$ is integral breadth due to strain effect. From equations (3) and (5), it was found that finite crystallite size induced peak broadening varies as function of $1/\cos\theta$ while strain broadening followed the nature of $\tan\theta$. In both cases, the reduction in crystallite size is predicted with at the higher diffraction angle. Furthermore, the root-mean-square value of microstrain  ($\epsilon_{rms}$) along the different crystallographic planes can also be estimated using Stocks-Wilson relations:
\begin{equation}
  \epsilon_{rms} = \sqrt{\frac{2}{\pi}}~{\epsilon}
\end{equation}
\begin{figure}[!ht]
  \begin{center}
\includegraphics[trim=0.3cm 0cm 0cm 0cm, clip=true, totalheight=0.6\textheight, angle=0]{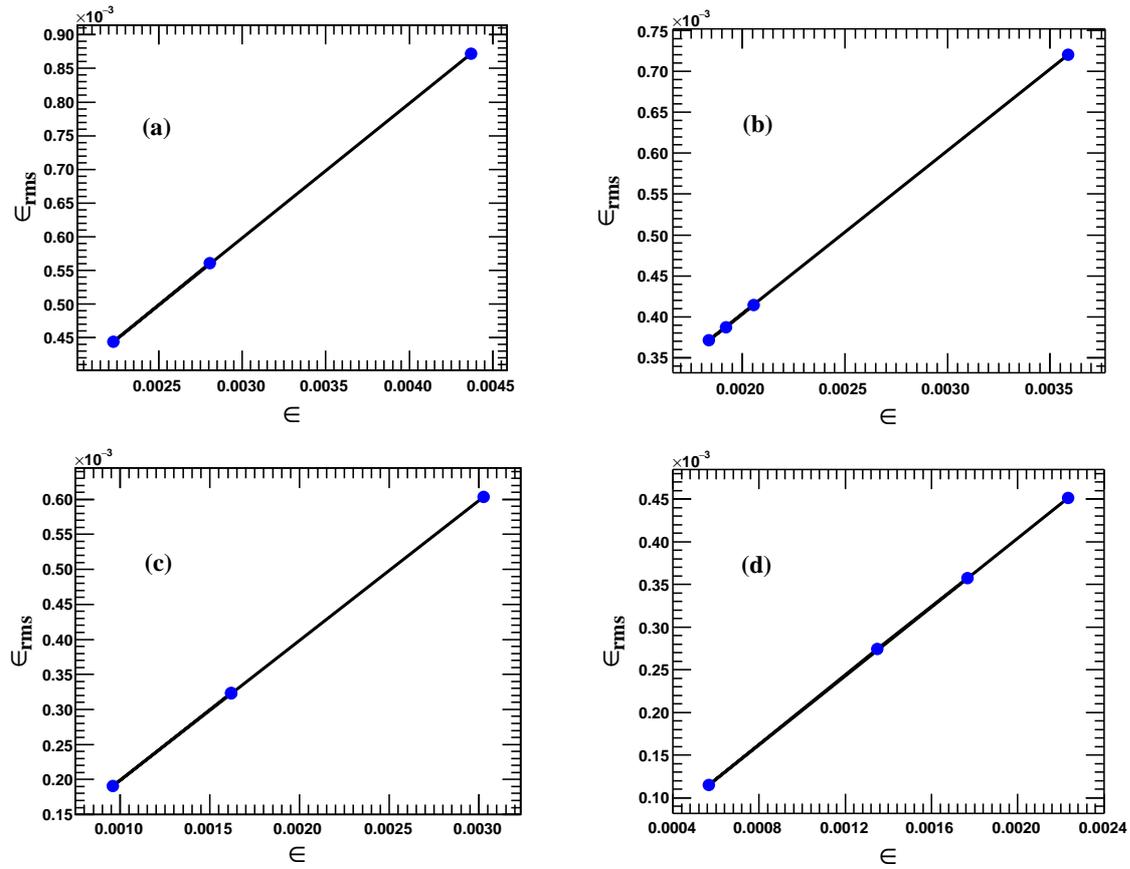}
 \caption{The plots of $\epsilon_{rms}$ versus $\epsilon$ of (a) 50, (b) 100, (c) 300 and (d) 500 nm KBr films.}
     \label{AFM_2d_3d}
    \end{center}
  \end{figure}

The plots of $\epsilon_{rms}$ verses $\epsilon$ are shown in Figure 3. Ideally, the data points should lie on straight line by making an angle of $45^{o}$ with abscissa. From Figure 3, it is evident that the root mean square strain linearly varies with microstrain which indicates that there is no discrepancy associated with the crystallographic direction of lattice planes. The results obtained from the Scherrer-Wilson relation are given in Table 3.

\subsubsection{Williamson-Hall Method:}
The Williamson-Hall method is a simple visualization of order dependence peak broadening. In this method, it is assumed that the size and strain contribution to the line broadening are mutually independent to each other and closely approximated by Cauchy's function and corresponding integral breadths are linearly additive: 
\begin{equation}
  \beta_{hkl} = \beta_{D} + \beta_{S}
\end{equation}  
Putting the value of $\beta_{D}$ and $\beta_{S}$ from equation (3) and (5), we get 
\begin{equation}
  \beta_{hkl} = \frac{K \lambda }{D \cos\theta} + 4\epsilon \tan\theta 
\end{equation}
By rearranging equation (8)
\begin{equation}
  \beta_{hkl}\cos\theta = \frac{K \lambda }{D} + 4\epsilon \sin\theta 
\end{equation}

\begin{figure}[!ht]
  \begin{flushleft}
\includegraphics[trim=0.3cm 0cm 0cm 5.5cm, clip=true, totalheight=0.25\textheight, angle=0]{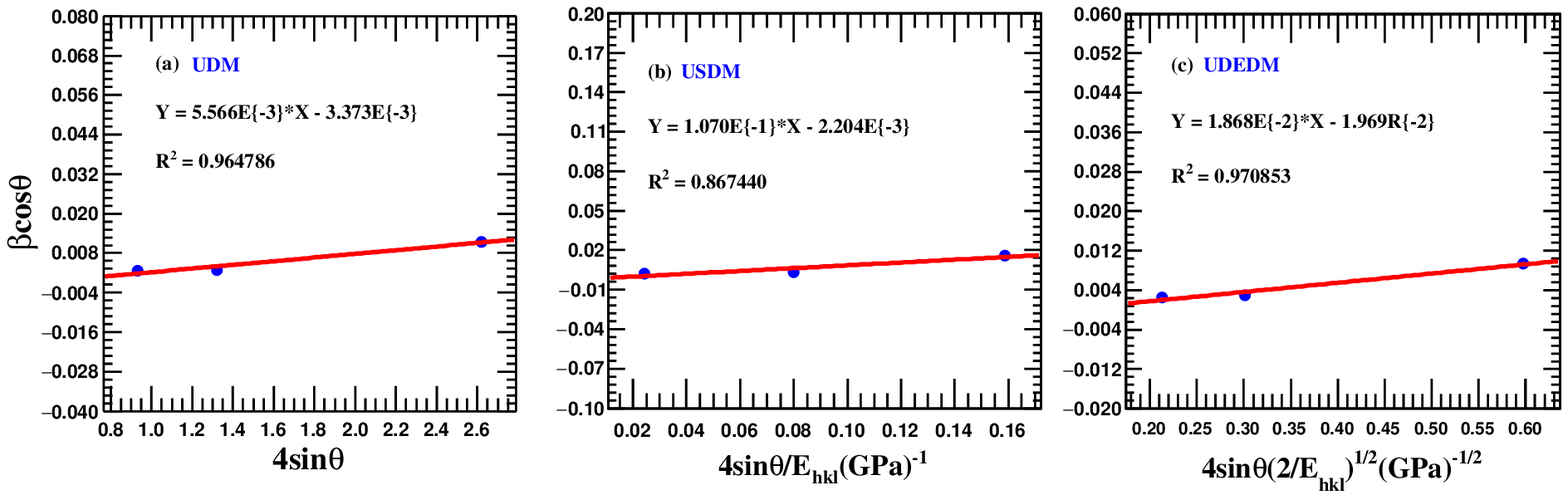}
\includegraphics[trim=0.5cm 0cm 0cm 5.7cm, clip=true, totalheight=0.26\textheight, angle=0]{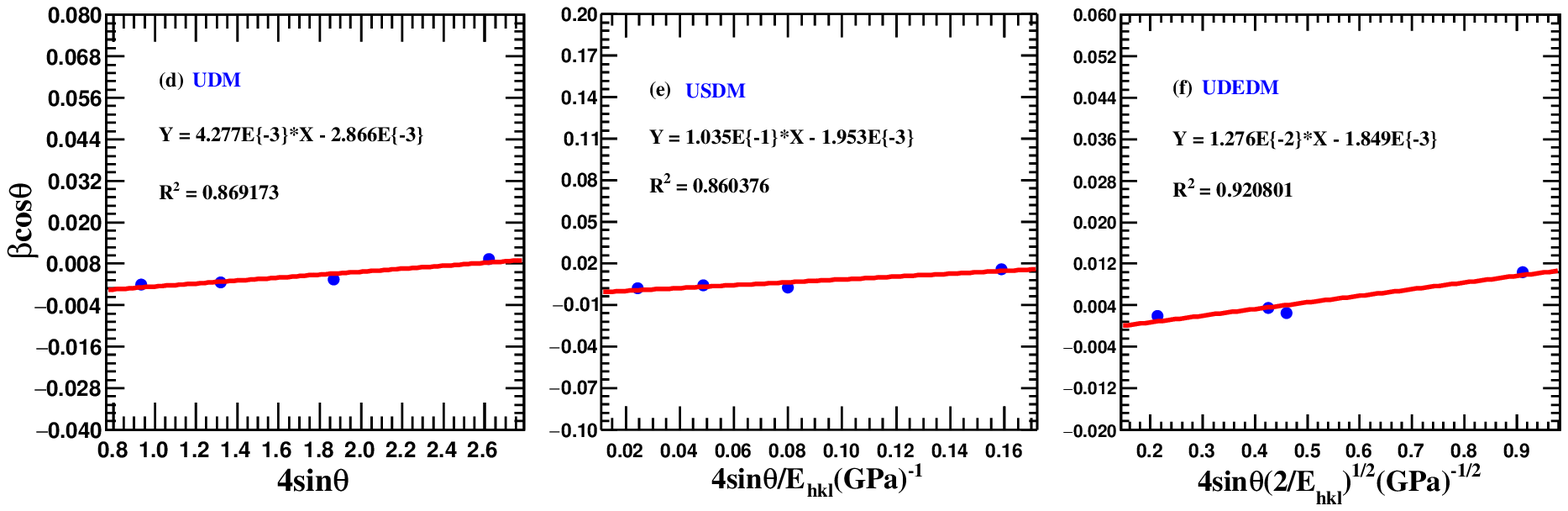}
 \caption{Williamson-Hall (a), (d) UDM, (b), (e) USDM and (c), (f) UDEDM plots of 50 (top panel) and 100 nm (bottom panel) thick KBr film.}
     \label{AFM_2d_3d}
    \end{flushleft}
  \end{figure}

After, plotting the values of  $\beta_{hkl}\cos\theta$ as a function of $4\sin\theta$, the intercept on vertical axis is a measurement of the volume of the domains which are diffracted coherently and slope gives ``effective" strain ~\cite{Hall}. Equation (9) represents the Uniform Deformation Model (UDM) of W-H method, in which strain is assumed to be uniform in all crystallographic direction ~\cite{Triloki}. Thus isotropic nature of the crystal is considered where material properties are assumed to be independent of the direction along which they are measured. This presumption is doubtful and therefore, in the other variant of W-H method more physical approach is adopted by considering an anisotropic magnitude of strain. The uniform deformation model of 50, 100, 300 and 500 nm KBr films is shown in Figure 4(a), 4(d), 5(a) and 5(d) respectively. 

In the Uniform stress deformation model (USDM), the isotropic stress $(\sigma)$ is assumed as a source of an anisotropic microstain. In this case, the isotropic microstrain  in UDM is replaced by $\epsilon_{hkl}$= $\sigma$/$E_{hkl}$, where $E_{hkl}$ is the Young's modulus in the direction perpendicular to crystallographic planes and W-H equation has the form

\begin{equation}
  \beta_{hkl}\cos\theta = \frac{K \lambda }{D} + \frac{4\sigma sin\theta}{E_{hkl}} 
\end{equation}

\begin{figure}[!ht]
\begin{flushleft}
 
\includegraphics[trim=0.3cm 0cm 0cm 5.5cm, clip=true, totalheight=0.26\textheight, angle=0]{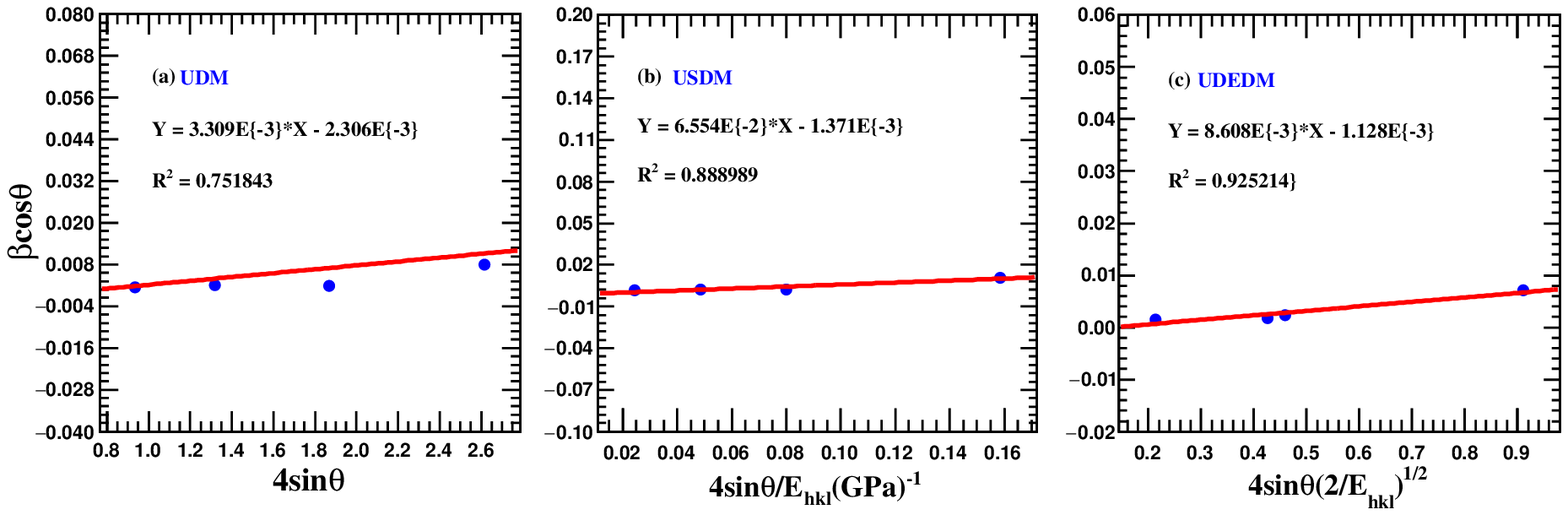}
\includegraphics[trim=0.3cm 0cm 0cm 5.5cm, clip=true, totalheight=0.25\textheight, angle=0]{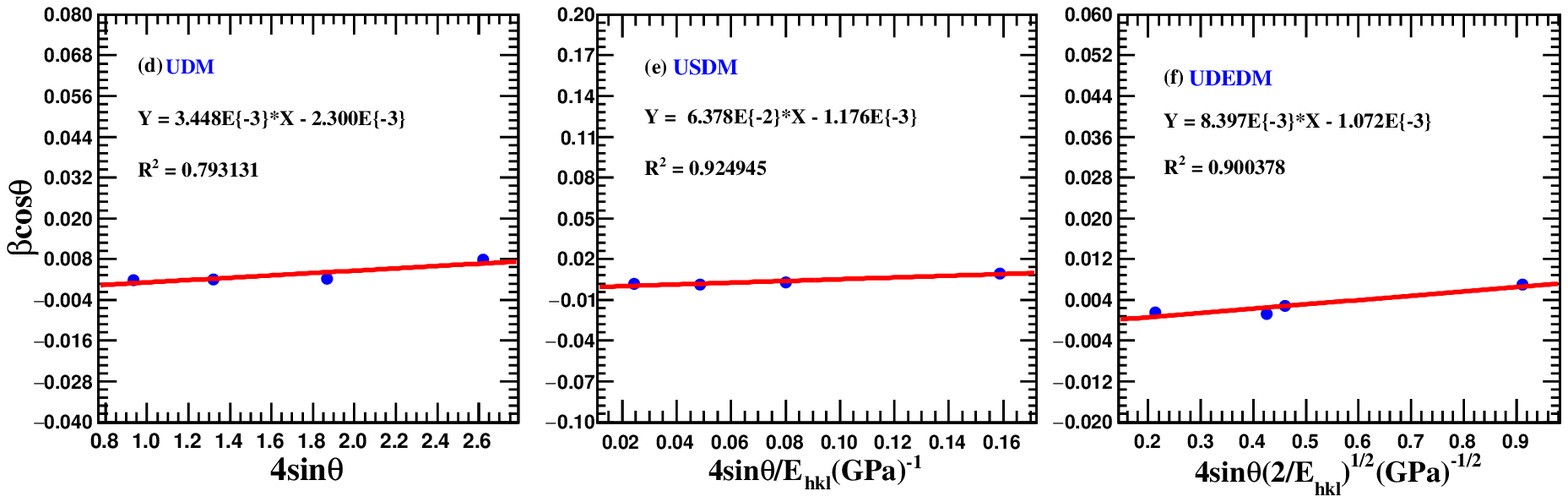}
 \caption{Williamson-Hall (a), (d) UDM, (b), (e) USDM and (c), (f) UDEDM plots of 300 (top panel) and 500 nm (bottom panel) KBr film.}
     \label{AFM_2d_3d}
    \end{flushleft}
  \end{figure}
In equation (10), an  anisotropic nature of Young's modulus is taken into account and the linear extrapolation of the plot drawn between $\beta_{hkl}\cos\theta$ and $4\sin\theta/E_{hkl}$ gives the value of uniform  stress, shown in Figure 4(b), 4(d), 5(b), 5(d). The strain can be calculated if the values of $E_{hkl}$ is known. For a cubic crystal Young's modulus, $E_{hkl}$ in the direction of the unit vector $l_{i}$ is given by the relation: 

\begin{equation}
\frac{1}{E_{hkl}} = s_{11}-2\big(s_{11}-s_{12}-\frac{1}{2}{s_{44}}\big)\big(l_{1}^{2}l_{2}^{2}+l_{2}^{2}l_{3}^{2}+l_{3}^{2}l_{1}^{2}\big)
\end{equation}

where, $s_{11}$, $s_{12}$ and $s_{44}$ are the elastic compliances of KBr. The explicit general equations which relate the elastic compliances $(s_{ij})$ to the stiffness $(c_{ij})$ are as follows ~\cite{BooK1}:

\begin{equation}
s_{11} = \frac{(c_{11} + c_{12})}{(c_{11} - c_{12})(c_{11} + 2 c_{12})}
 \end{equation}

\begin{equation}
s_{12} = \frac{- c_{12}}{(c_{11} - c_{12})(c_{11} + 2 c_{12})}
 \end{equation}

\begin{equation}
s_{44} = \frac{1}{c_{44}}
 \end{equation}
Where, corresponding value of $c_{11}$, $c_{12}$ and $c_{44}$ at room temperature are $3. 46\times10^{11}dynes/cm^{2}$, $0. 58\times10^{11}dynes/cm^{2}$, $0.505\times10^{11}dynes/cm^{2}$ respectively ~\cite{galt}. 

In the Uniform Deformation Energy Density Model (UDEDM), the  density of deformation energy is assumed to be uniform in all crystallographic direction and considered as a cause of anisotropic lattice strain. Moreover, proportionality constant ($E_{hkl}$) associated with the stress-strain relation is no longer independent when the strain energy density is involved. According to Hook's law, the microstrain is a function of square root of energy density $(u)$, $\epsilon_{hkl}$ = $(2u/E_{hkl})^{1/2}$. Therefore equation (10) can be modified to the following form: 
\begin{equation}
  \beta_{hkl}\cos\theta = \frac{K \lambda }{D} + {4\sin\theta}{\bigg(\frac{2u}{E_{hkl}}\bigg)^{1/2}} 
\end{equation}
Here $u$ is the energy per unit volume. The uniform deformation energy can be estimated from the slope of line plotted between  $\beta_{hkl}\cos\theta$ and $4\sin\theta(2u/E_{hkl})^{1/2}$. Further, similar to the previous cases, crystallite size can be calculated from intercept on vertical axis. Figure 4(c), 4(f), 5(c) and 5(f) represent the uniform deformation energy density models of KBr films. The value of Young's modulus $(E_{hkl})$ has been calculated from equation (11) and found to be about 33.3069 GPa for (200) and (400) crystallographic planes and 13.8539 GPa for (220) and (440) planes. All the structural parameters, evaluated from W-H models, are reported in Table 2.

\begin{table*}[tp]
\setlength{\tabcolsep}{3pt}
\renewcommand{\arraystretch}{1.5}
\caption{Structural profile of KBr films of different thickness using Whole patterns fitting Methods.}
\begin{flushleft}
\scriptsize {\begin{tabular}{l*{13}rrr}
 \toprule
  Samples &&&& \multicolumn{3}{c}{ Williamson-Hall Method } &&&& \multicolumn{4}{c}{ Size-Strain Plot Method} \\ 
\cmidrule(r){2-10}
&\multicolumn{2}{c}{UDM}   &\multicolumn{3}{c}{USDM}  &\multicolumn{4}{c}{UDEDM}   && &  &  \\
\cmidrule(r){2-3}
\cmidrule(r){4-6}
\cmidrule(r){7-10}
         & $D$ & $\epsilon\times10^{-3}$ &  $D$    &  $\epsilon\times10^{-3}$ & $\sigma\times10^{-2}$ &  $D$  & $\epsilon\times10^{-3}$ & $\sigma\times10^{-2}$  & $u$ & $D$ & $\epsilon\times10^{-3}$ & $\sigma\times10^{-2}$ &  \\
  & $(nm)$ & $no~unit$ &  $(nm)$ &  $no~unit$ & $ MPa$ & $(nm)$ & $ no~unit$ & $MPa$ & $KJm^{-3}$ & $(nm)$ & $ no~unit$ & $MPa$ \\
\midrule
 50 nm   & 41.11   & 5.57   & 62.92   & 4.495    & 10.70 & 70.41 &5.42 &12.90 &349.27& 66.42& 3.33 & 7.92\\
 100 nm  & 48.36   & 4.28  & 71.02  & 3.77 & 10.35&74.96&3.44&9.46&162.88&76.47&2.58&7.08\\
 300 nm  & 53.98   &3.52   & 101.12  & 2.63 &7.21 & 122.86&2.32&6.38&74.10&110.39&1.95&5.36\\
 500 nm  & 60.18   & 3.45  & 117.90 & 2.32& 6.38 & 129.38&2.27&6.22&70.51&130.53&1.55&4.27\\
    \bottomrule
    \bottomrule
 \end{tabular}}
\end{flushleft}
\label{table:nonlin}
\end{table*}

\subsubsection{Size-Strain Plot Method}
\begin{figure}[!ht]
  \begin{center}
\includegraphics[scale=0.88]{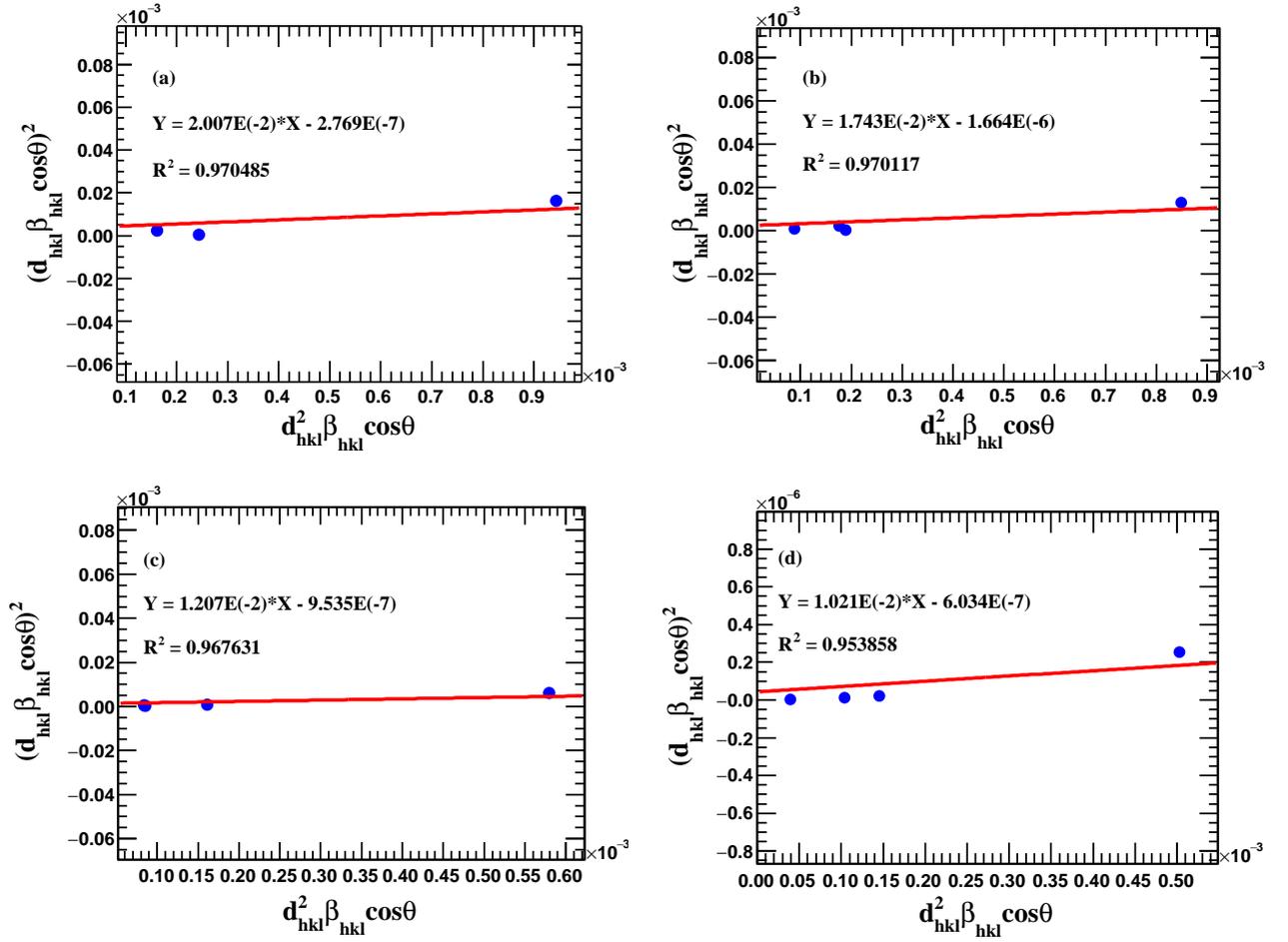}
 \caption{The Size-Strain plots of (a) 50, (b) 100, (c) 300 and (d) 500 nm KBr films.}
     \label{size_strain}
    \end{center}
  \end{figure}
W-H method showed that line broadening in XRD peaks is essentially isotropic and least square fitting of data points gives positive value slope and non-zero intercept. However, in the cases of isotropic line broadening, a better estimation of the size and strain parameters can be obtained from the ‘Size-Strain Plot’ method (SSP), which has an advantage over W-H plots that less importance is given to the reflections occur at high angles where the precision is usually lower \cite{Tagliente}. In this method, it is assumed that the ‘strain profile’ is explained by a Gaussian function and the ‘crystallite size’ profile by a Cauchy's function and given by following equation:
\begin{equation}
\big(d_{hkl}~\beta_{hkl}\cos\theta\big)^{2}= {\frac{1}{V_{s}}}\big(d_{hkl}^{2}~\beta_{hkl}\cos\theta\big) + \Big(\frac{\epsilon_{hkl}}{2}\Big)^{2}
 \end{equation}
where $d_{hkl}$ is the lattice spacing between the adjacent $(hkl)$ planes, $V_{s}$ is the apparent volume weighted average size and $\epsilon_{hkl}$ is the apparent microstrain. For spherical crystallites the volume average true size is given by $<D_{v}>$ =  4/3$V_{s}$. The plots of $(d_{hkl}\beta_{hkl}\cos\theta)^{2}$ Vs. $(d_{hkl}^{2}\beta_{hkl}\cos\theta)$ of (a) 50 nm, (b) 100 nm, (c) 300 nm, (d) 500 nm KBr films are shown in Figure 6. The crystallite size ($V_{s}$) and apparent microstrain have been estimated from the slope and the intercept of the linear extrapolated data, respectively and shown in Table 2.

\subsubsection{Single line Voigt Method}
\begin{figure}[!ht]
  \begin{center}
\includegraphics[trim=1.5cm 0.5cm 2.5cm 1.5cm, clip=true, totalheight=0.30\textheight, angle=0]{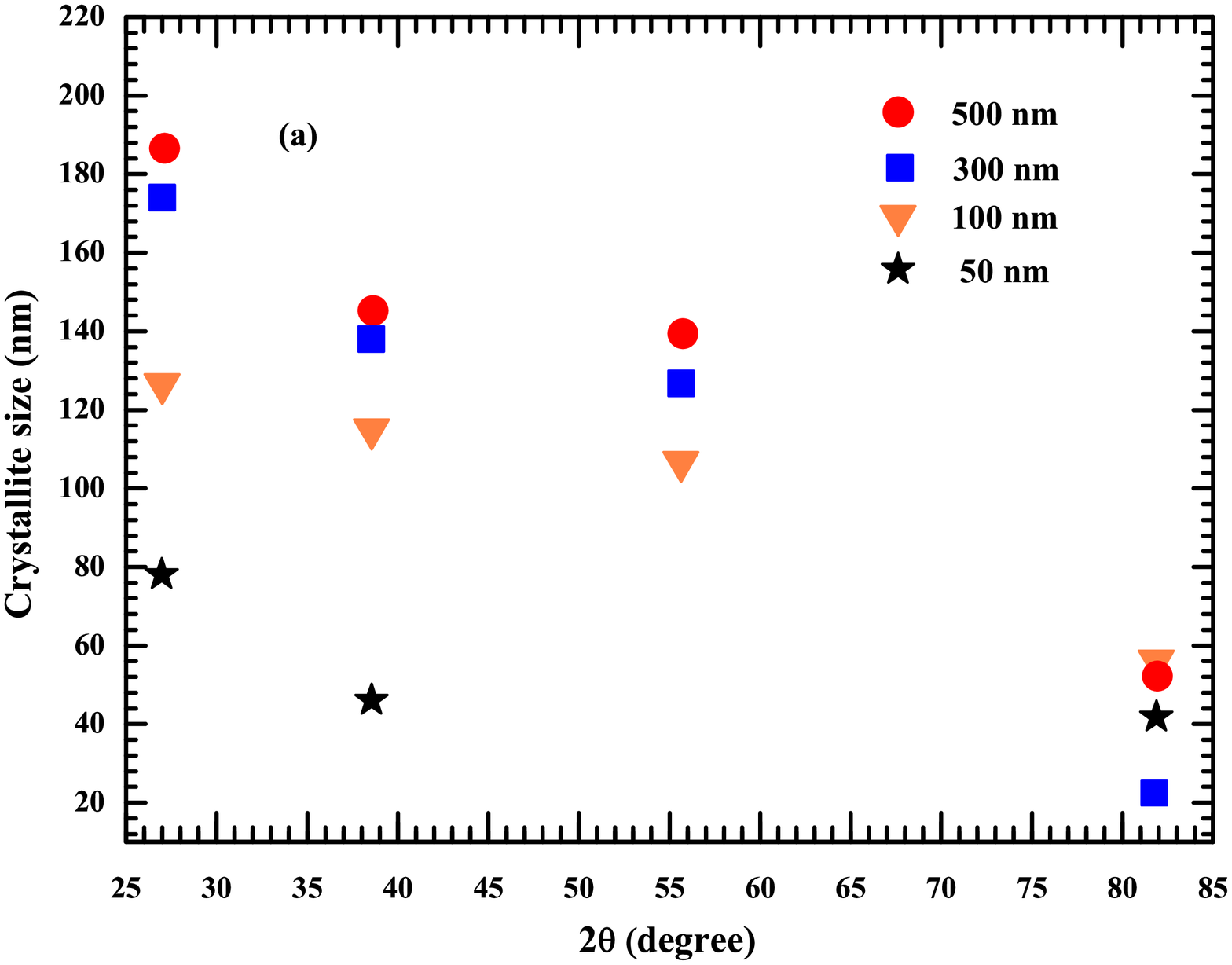}
\includegraphics[trim=1.0cm 0.5cm 2.5cm 1.5cm, clip=true, totalheight=0.30\textheight, angle=0]{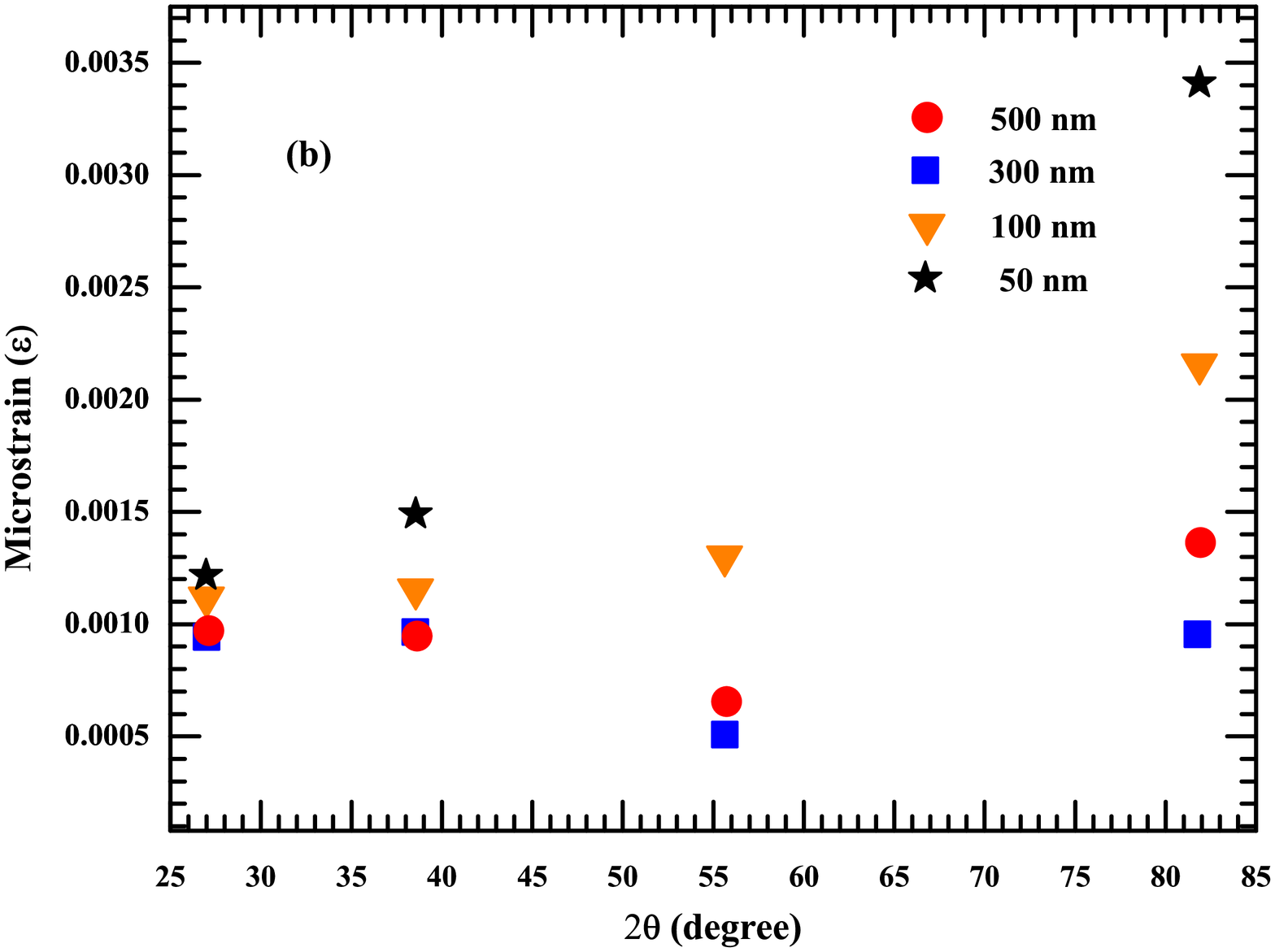}
 \caption{ (a) Crystallite size and (b) Microstrain of KBr films calculated by single line Voigt method.}
     \label{AFM_2d_3d}
    \end{center}
  \end{figure}

The W-H method makes use of at least three reflection from which an integral breadth can be determined, while the Voigt function provides a rapid and powerful single line method for the analysis of diffracted peak breadths in terms of crystallite size and lattice strain. The line profile due to finite crystallite size causes a Cauchy-type broadening while strain profile tends to Gaussian and the Voigt function can be represented as the convolution of these two profiles (Cauchy's and Gaussian), each having the same origin. Based on these observations the Cauchy and Gauss components of a single diffraction peak can be interpreted to the individual effect of size and strain induced broadening. If, the experimental broadened profile ($h$), pure sample profile ($f$) and instrumental profile $(g)$ are Voigtian then
\begin{equation}
 h_{C} = g_{C}*f_{C}
\end{equation} 
\begin{equation}
h_{G} = g_{G}*f_{G}
 \end{equation}
Where subscript C and G denote the Cauchy and Gaussian components of respective Voigt profiles. In this case, Cauchy's and Gaussian components can be extracted from sample broadened profile after decomposition of instrumental broadening.    
\begin{equation}
 \beta_{C}^{f} = \beta_{C}^{h} - \beta_{C}^{g}
\end{equation}
and
\begin{equation}
 (~\beta_{G}^{f}~)^{2} =(~\beta_{G}^{h}~)^{2} - (~\beta_{G}^{g}~)^{2} 
\end{equation}
Langford (1978)~\cite{Langford2} subsequently derived an explicit equation for the Voigt function and proposed that the fraction of the Cauchy and Gaussian components of sample profile can be easily obtained from the ratio of the FWHM $(2w)$ to the integral breadths $(\beta)$. Alternatively, in order to avoid graphical interpolation, Keijser $et~al.$ \cite{Keijser} showed that the approximate values of the constituent breadths can be more conveniently obtained by following empirical formula:

\begin{equation}
 \frac{\beta_{c}}{\beta} = (a_{0}+a_{1}\psi+a_{2}\psi^{2});
 \end{equation}
and
\begin{equation}
  \frac{\beta_{g}}{\beta} = b_{0}+ b_{1/2}\bigg(\psi-\frac{2}{\pi}\bigg)^{1/2}+b_{1}\psi+b_{2}\psi^{2}
\end{equation}

   Where, $\psi = 2w/\beta$, $a_{0}$ = 2.0207, $a_{1}$ = -0.4803, $a_{2}$= -1.7756, $b_{1/2}$ = 1.4187, $b_{0}$ = 0.6420, $b_{1}$ = -2.2043 and
   $b_{2}$ = 1.8706.
In a single line Voigt analysis apparent crystallite size is given by 
\begin{equation}
  D= \frac{K \lambda }{ \beta_{C}^{f} \cos\theta} 
\end{equation}
and the heterogeneous microstrain $\epsilon$ by 
\begin{equation}
  \epsilon = {\frac{\beta_{G}^{f}}{4}}{\cot\theta}  
\end{equation} 

   Where, integral breadth $\beta$ is measured on $2\theta$ scale, $\lambda$ is the wavelength of $CuK_{\alpha_{1}}$ and $\theta$ is the angular position of the diffracted peak. Figure 7(a) and (b) illustrated the results obtained from the single line Voigt approximation. The peak analysis reveals that the values of crystallite sizes vary significantly from peak to peak and larger domain sizes corresponds to (200) crystallographic plane. However, relatively smaller crystallite sizes are extracted from other peak positions which indicates the increment in a peak tail (Cauchy's component) with a decreasing d-spacing. For (440) plane, the value of apparent domain sizes, is smallest and similar impression observed from Scherrer equation, which exhibit the larger orientational difference and dislocation density of corresponding plane. After the evaluation of microstrain using equation (14), it  is found that the  strain value is not increased monotonously with increasing diffraction angle and very small microstrain gradient resides within the crystallite of (200) planes of different film thickness. The smaller values of strain  correspond to (400) lattice plane of 300 nm and 500 nm films, while larger values obtained for (440) crystallographic plane of 50 nm film. The estimated value of crystallite size and microstrain, along each crystallographic plane, are shown in Table 3.

\begin{table*}[tp]
\setlength{\tabcolsep}{5pt}
\renewcommand{\arraystretch}{0.8}
\caption{Structural profile of KBr films using Single peak analysis methods.}
\begin{flushleft}
 {\begin{tabular}{l*{12}rrr}
\toprule
  Samples & $hkl$& \multicolumn{2}{c}{Scherrer Method} & \multicolumn{2}{c}{ Wilson Method}& \multicolumn{3}{c}{Voigt Method} & $TC_{(hkl)}$\\  
\cmidrule(r){2-2}
\cmidrule(r){3-4}
\cmidrule(r){5-6}
\cmidrule(r){7-9}
      & &$D$ & $\delta\times10^{-4}$    & $\epsilon\times10^{-3}$&   $\epsilon_{rms}\times10^{-4}$  & $D$ & $\epsilon\times10^{-3}$ & $\delta\times10^{-5}$&&\\
  & & (nm)  & (nm)$^{-2}$&  & & (nm) &  &(nm)$^{-2}$\\
 
\midrule
 50 nm      &200 & 53 & 3.56 & 2.80     & 5.59  & 78  & 1.22  & 16.44  &0.31\\
            &220 & 47 & 4.50 & 2.23     & 4.44  & 46  & 1.49  & 47.14  & 1.31\\
            &440 & 12 & 68.17  & 4.37   & 8.72  & 42  & 3.41  & 57.33  & 2.31\\

100 nm & 200 & 72    & 1.92  & 2.06  & 4.11  & 126   & 1.12  & 6.24     &1.26\\
       & 220 & 54    & 3.34  & 1.92  & 3.83  &115    & 1.15  & 7.55     &0.14\\
       &400  & 40    &6.11    & 1.84 & 3.66  &106    &1.30   & 8.79     &1.43\\
       &440  &14     &46.10   & 3.59 & 7.17  &56     &2.15   & 31.75    &1.16\\

300 nm &200  & 91    & 1.19  & 1.62   &3.23  & 174 & 0.94  & 3.30   &2.71 \\
       &220  & 64    & 2.37  & 1.62   &3.23  &138  & 0.96  & 5.24   &0.12 \\
       &400  &77     & 1.66  & 0.96   &1.91  &126  & 5.08  & 6.23   &1.06 \\
       &440  &17     & 32.59 & 3.02   &6.03  &22   & 0.95  & 196.88 &0.10 \\

500 nm & 200 & 80    &1.54  & 1.79   & 3.57  & 186  & 0.97 & 2.87    &1.87 \\
       & 220 & 72    &1.91  & 1.37   & 2.73  & 145  &0.95  & 4.73    &0.02 \\
       & 400 & 113   &0.77  & 0.566  & 1.15  & 139  &0.65  & 5.14    &1.95 \\
       & 440 & 18    &32.05 & 2.26   & 4.51   & 52  &1.36  & 36.70   &0.16 \\
    \bottomrule
    \bottomrule
 \end{tabular}}
\end{flushleft}
\label{table:nonlin}
\end{table*}

The aim of pattern deconvolution in this context is to obtain reliable estimates of line-profile parameters, particularly the position, height and area of each diffraction peak for the use in an analysis of structural imperfection. Figure 8(a) and (b) represents the comparative study of average crystallite size and microstain determined by different analytical technique. Among the whole pattern fitting methods, the SSP analysis gives the lowest value of the lattice strain and larger values of crystallite sizes, but the application of this method is limited to  isotropic line broadening. In the case of single line Voigt approximation, it has to be noted that the  crystallite size and microstrain are  strongly dependent on the ratio of FWHM to integral breadth and this ratio must be laid between Cauchy and Gaussian limit which restrain the application of this method. Concerning to other single line methods, the Scherrer method seems to underestimate the value of crystallite size and therefore over estimate the lattice strain and it is also ignore peak broadening resulting from other important factors such as inhomogeneous strain and instrumental effects. Same results can be derived from the Wilson relation. Since in both cases, the peak broadening attributed to the existence of either due to the finite crystallite size effect (Scherrer equation) or solely strain effect (Wilson's relation) and it is assumed that all the crystallites are quasi isotropic.

\begin{figure}[!ht]
  \begin{center}
\includegraphics[trim=1.5cm 0.5cm 2.5cm 1.5cm, clip=true, totalheight=0.28\textheight, angle=0]{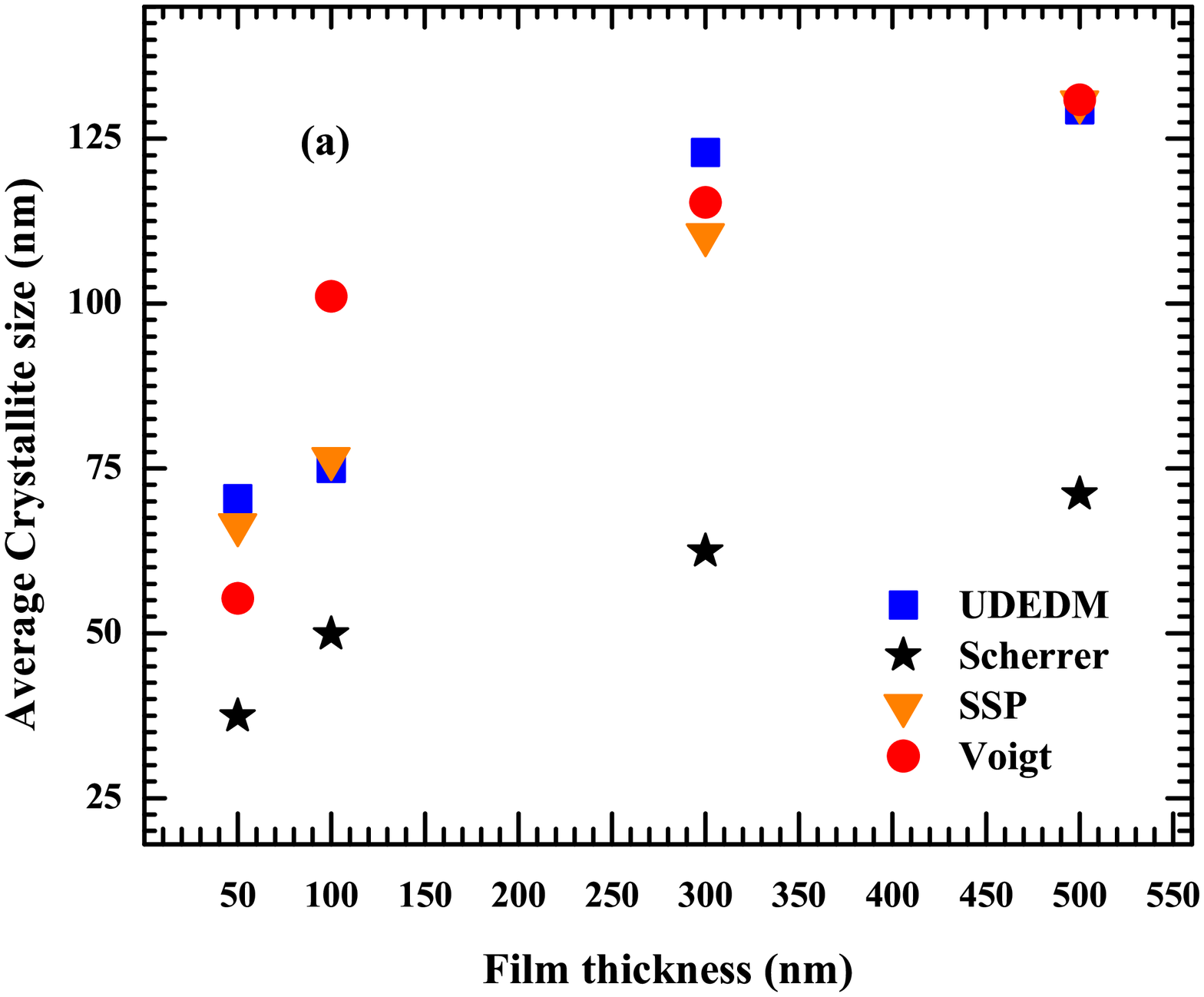}
\includegraphics[trim=1.5cm 0.5cm 2.5cm 1.5cm, clip=true, totalheight=0.28\textheight, angle=0]{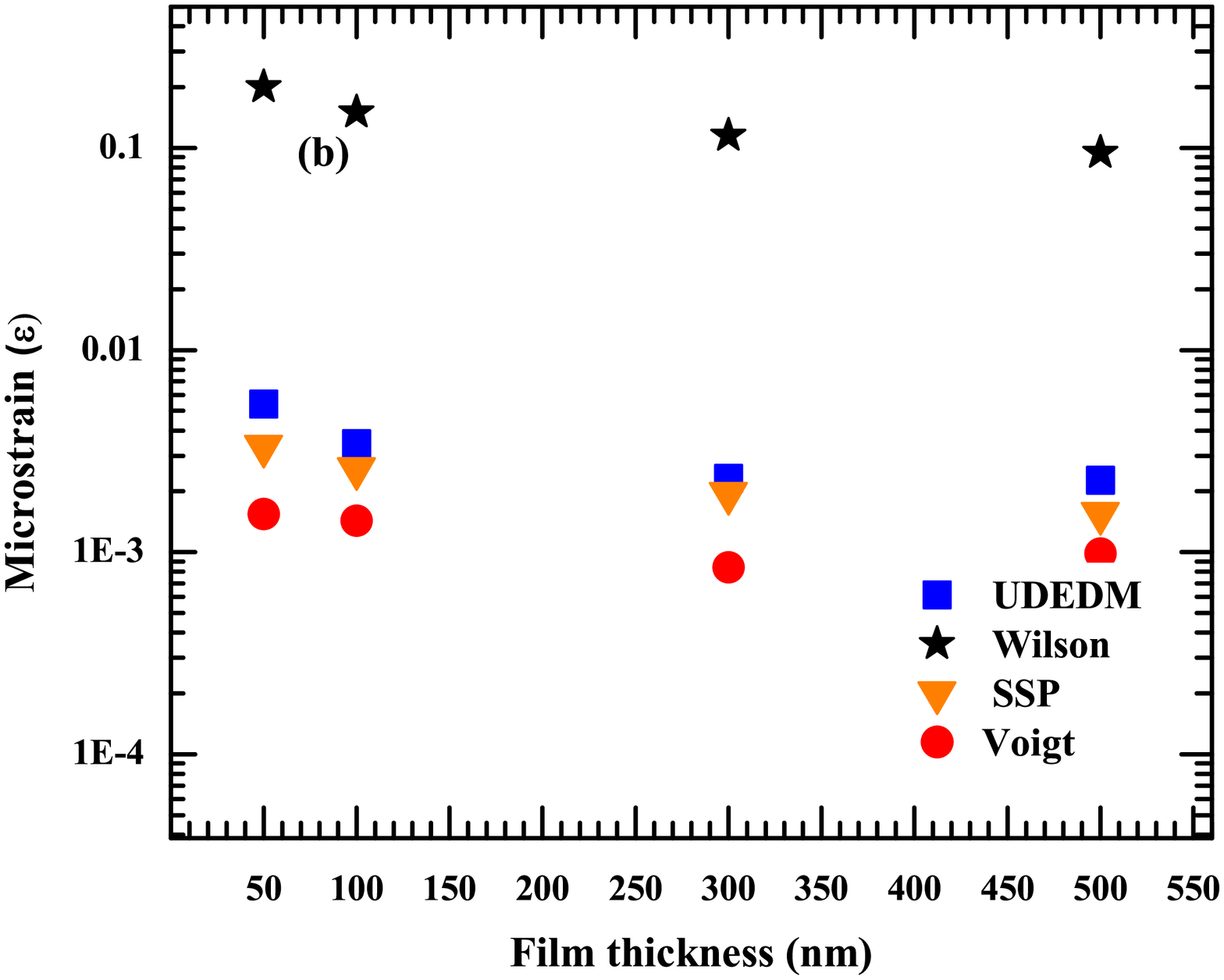}
 \caption{The average values of (a) Crystallite size and (b) Microstrain KBr films.}
     \label{AFM_2d_3d}
    \end{center}
  \end{figure}

The results derived from UDM, USDM and UDEDM approximation of W-H  predict a different value of crystallite size and microstrain. For a given sample, W-H plots may be plotted using equations (16) to (24) and the most suitable approach is the one for which experimental data points are best fitted. A comparison of these three evaluation procedures for the size-strain evaluation of KBr film is summarized in Table 3. From the analysis of results obtained from different W-H plots, it is evident that  the magnitude of lattice distortion decreases and therefore the size of coherently scattering domain increases with increasing film thickness. This observation is also supported by the results extracted from the other analytical methods. Further, from Figure 6 and 7, it is observed that the scattering of experimental data points away from the linear expression is lesser in the case of UDEDM approach as compared to UDM and USDM W-H models. It can be noted that the values of the average crystallite size obtained from the UDEDM approach are in good agreement with the results of the SSP method. Thus, it may be concluded that in present case UDEDM is being more realistic approximation than the other methods. As far authors are concerned, a detailed study using X-ray diffraction line profile of KBr thin films is not reported yet. This study throws some more light and reveals the importance of analytical models in the determination of crystallite size and lattice distortion of thin films. We suggest that UDEDM approximation is the best approach for the evaluation of the crystallite size and anisotropic value of microstrain.

 \section{Conclusion} 
In this work, we have described and exploited different analytical techniques for investigating the X-ray diffraction profile of KBr films. Emphasis is laid on the extensive study of peak broadening in terms of integral breadth rather than full width at half maximum. Homogeneous as well as heterogeneous strain is observed in the KBr films and its value decreases with increasing film thickness. The Cauchy and Gaussian component of the Voigt function have been determined for along all diffraction peak. The size and strain contribution is investigated by Williamson-Hall Uniform Deformation, Uniform Stress Deformation and Uniform Deformation Energy Density approximation. Among them Uniform Deformation Energy Density model is most appropriate approach and than the observed results are compared with other methods. The average value of crystallite size, estimated from Voigt method, are larger than those from Williamson-Hall and Size-Stress Plot method, but almost follow similar pattern $i. e.$ increases with increasing film thickness. The reduction in crystallite size is observed as column length becomes sharper for the thinner film. All the models yield different values of crystallite size and lattice strain, it may be due to the anisotropic nature of line broadening and different assumption about the peak shape. The Scherrer method and single line Voigt analysis reveal that the crystallite size decreases while microstrain increases at the higher diffraction angle which is supported by the TC$_{hkl}$ and dislocation density calculation.

 \section{Acknowledgment}
  This work was partially supported by the Department of Science and Technology (DST)(Grant no. SR/S2/HEP-19/20060), the Council of Scientific and Industrial Research (CSIR) (Grant no.03(1992)/07EMR-II) and the Indian Space Research Organization (ISRO). R. Rai acknowledge University Grant Commission (UGC), New Delhi, India for providing financial support.


\begin{thebibliography}{00}

      \bibitem{SUMER}U. Feldman, W.E.Behring,  W. Curdt et al., Astrophysical Journal Supplement, 113, 195 (1997). 

      \bibitem{PHEBUS}K. Yoshioka, K. Hikosaka, G. Murakami et al., Advances in Space Research 41, 1392 (2008).        
	
      \bibitem{over}H. W. Moos, W. C. Cash, L. L. Cowie et al., Astrophysical Journal 538, L1 (2000).

      \bibitem{E. Figueroa}E. Figueroa-Feliciano, Progress in Particle and Nuclear Physics 66, 661 (2011).
 
      \bibitem{ST Coyle}S. T. Coyle, B. Shamoun, M. Yu, J. Maldonado et al., Journal of Vacuum Science \& Technology B 22, 501 (2004).

      \bibitem{Sergei} Sergei A. Aseyev, Peter M. Weber, Anatoli A. Ischenko, Journal of Analytical Sciences, Methods and Instrumentation, 3, 30 (2013).

      \bibitem{J. Workman} J. Workman, A. Maksimchuk, X. Liu et al., PHYSICAL REVIEW LETTERS, 75, 2324 (1995).

      \bibitem{D. Mormann}D. M ormann, A. Breskin, R. Chechik, et al., Nuclear Instruments and Methods in Physics Research A 478, 230 (2002).

      \bibitem{Daisuke}Daisuke Totsuka, Takayuki Yanagida, Yutaka Fujimoto et al., Optical Materials, 34, 1087, (2012).

      \bibitem{Wei} Wei Zhao,Goran Ristic, J. A. Rowlands, Medical Physics, 31, 2594 (2004). 

     \bibitem{F. Garibaldi}F. Garibaldi,  E. Cisbani, F. Cusanno et al. Nuclear Instruments and Methods in Physics Research A, 525, 263 (2004).
 
     \bibitem{OHW}Oswald H. W. Siegmund, E. Everman, J. V. Vallerga, et al., Applied Optics, 26, 3607 (1987).

     \bibitem{A. Breskin1996}A. Breskin, A. Buzulutskov, R. Chechik et al., Appied Physics Letters, 69, 1008 (1996).
	
      \bibitem{Oswald}Oswald H. W. Siegmund, D. E. Everman, J. V. Vallerga et. al., Applied Optics, 27, 1568 (1988).

	 \bibitem{Tremsin}A.S. Tremsin, O.H.W. Siegmund, Nuclear Instruments and Methods in Physics Research A 442, 337 (2000).
	 
	 \bibitem{Weidong}Weidong He,Subramanian Vilayurganapathy, Alan G. Joly, et al., Applied Physics Letters 102, 71604 (2013).

	  \bibitem{Rai}R. Rai, Triloki, N. Ghosh, et al., Nuclear Instruments and Methods in Physics Research A 787, 125 (2015).

	\bibitem{Luna}M. Luna, F. Rieutord, N. A. Melman et. al., J. Phys. Chem. A, 102, 6793 (1998).

	  \bibitem{Nitti}M.A. Nitti, N. Cioffi, E. Nappi, et al., Nuclear Instruments and Methods in Physics Research A 493, 16 (2002).

	 \bibitem{Buzulutskov}A. Buzulutskov, A. Breskin, R.Chechik, Nuclear Instruments and Methods in Physics Research A 366, 410 (1995).

	\bibitem{BKSingh}B.K. Singh, E. Shefer, A. Breskin, Nuclear Instruments and Methods in Physics Research A 454, 364 (2000).

	 \bibitem{Singh1}B. K. Singh, M.A. Nitti, A. Valentini et al., Nuclear Instruments and Methods A 582, 651 (2007).

	\bibitem{Shefer}E. Shefer, A. Breskin, T. Boutboul, R. Chechik, et al.,Journal of Applied Physics 92, 4758 (2002).
	  

	\bibitem{BooK}Mario Birkhoz, Thin Film Analysis by X-Ray Scattering, Wiley Library (2006).

       
        \bibitem{T}T. Ung$\acute{a}$r, Scripta Materialia, 51, 777 (2004).

  
	\bibitem{Kisielowski}C. Kisielowski, J. Kruger, S. Ruvimov et al., PHYSICAL REVIEW B, 54, 17745 (1996).


        \bibitem{TAMÁS UNGÁR} T. Ung$\acute{a}$r, L. Balogh and G. Rib$\acute{a}$rik, Metall.  Mater.  Trans.  A, 41, 1202 (2010).


	\bibitem{Warren}B.E. Warren, X-ray Diffraction Addison Wesley, Reading, (1969).

	\bibitem{Langford}J.I. Langford, et al., Aust. J. Phys. 41, 173 (1988).

	\bibitem{Prince}E. Prince, J.K. Stalick, Accuracy in Powder Diffraction II, NIST Special Publication, 597, (1992).
        \bibitem{Hall} W.H. Hall, Proc. Phys. Soc. A,  62741 (1949).
      
        \bibitem{Barret} C. Barret, T.B. Massalki, Structure of Metals, Pergamon, Oxford (1980).  

        \bibitem{Cullity} Cullity, D. B. Elements of X-ray Diffraction , 2nd ed.; Addision-Wesley: Reading, MA (1978).

       \bibitem{Scherrer}P. Scherrer, Nachrichten von der Gesellschaft der Wissenschaften zu Gottingen26 (1918) 98.

       \bibitem{YU}Yu Rosenberg, V Sh Machavriani, A Voronel et al., J. Phys.: Condens, Matter 12, 8081 (2000). 

       \bibitem{Williamson}G. K. Williamson, R. E. Smallman, Philosophical Magazine, 1:1, 34 (1956).
 
	\bibitem{Stokes}A.R. Stokes, A.J.C. Wilson, Proc. R. Soc. London 56, 174 (1944).

	
  	

        \bibitem{Triloki}Triloki, P. Garg, R. Rai et al., Nuclear Instruments and Methods A 736, 128-134 (2014) and references therein.

 	 \bibitem{BooK1}J. F. Nye, PHYSICAL PROPERTIES OF CRYSTALS, Oxford University Press (2006).

	
	\bibitem{galt} J. K. Galt, Physical Reviews, 73, 1460 (1949).

\bibitem{Tagliente} M.A. Tagliente, M. Massaro, Nuclear Instruments and Methods in Physics Research B 266, 1055 (2008).

	\bibitem{Langford2} J. I. Langford,  Journal of Applied Crystallography, 11, 10 (1978).

	\bibitem{Keijser} Th. H. de Keijser, J. I. Langford, E. J. Mittemeijer et al., Journal of Applied Crystallography, 15, 308 (1982).


	\end{thebibliography}
\end{document}